\documentstyle[psfig]{mn}

\title[The absorption line spectrum of the QSO J2233$-$606]{The absorption line spectrum of J2233$-$606}
\author[P.~J.~Outram et al.]{P.~J.~Outram$^1$, B.~J.~Boyle$^2$, R.~F.~Carswell$^1$, P.~C.~Hewett$^1$,
\newauthor
R.~E.~Williams$^3$, R.~P.~Norris$^4$\\
$^1$ Institute of Astronomy, Madingley Road, Cambridge CB3 0HA\\
$^2$ Anglo-Australian Observatory, PO Box 296, Epping, 
NSW 2121, Australia\\
$^3$ Space Telescope Science Institute, 3700 San Martin Drive, Baltimore, MD
21218  USA\\
$^4$ Australia Telescope National Facility, Epping, NSW
2121, Australia}

\begin{document}
\maketitle
\begin{abstract} 
We report on high resolution observations ($R=35000$) of the Hubble
Deep Field South QSO J2233-606 obtained with the University College 
London Echelle Spectrograph at the AAT.  We present spectral
data and an associated absorption line list for the wavelength 
region 3530$ < \lambda <$ 4390\AA. The data has a mean 
signal-to-noise ratio in the continuum of approximately 15 per 0.05\AA\ 
resolution element at $\lambda >$ 3700\AA.
\end{abstract}
\begin{keywords}
 quasars: absorption line systems -- quasars: individual : J2233-606
\end{keywords}

\maketitle

\section{Introduction}

 The Hubble Deep Field \cite{wil96} provided unprecedented
information on the nature of the Universe at high redshifts (e.g.
Abraham et al. 1996 , Steidel et al. 1996, Madau et al. 1996). A similar
study of a southern field in the HST continuous viewing zone has just
been completed. A primary goal of
the Hubble Deep Field South (HDF-S) project will be to further our
understanding of the relationship between galaxies and quasar
absorption line systems and the HST field will be centred on a bright
high--redshift quasar.  The position of the HDF-S field was determined
by the location of the bright ($B=17.5$) $z=2.24$ QSO J2233--606,
discovered by one of us (PCH) and M J Irwin using APM scans of a UKST 
objective--prism plate, with confirmation of the object by the AAT 
\cite{boy97}.

The main reasons for centring the HDF-S on a
high redshift quasar are to compare the absorption line spectrum of the
quasar with the redshifts of the galaxies identified in the field, and
to probe the quasar environment. 

Initial spectral observations of J2233--606 have already been reported by Sealey, Webb \& Drinkwater \shortcite{sdw98} and Savaglio \shortcite{sav98}. Sealey  et al.\ \shortcite{sdw98} combined observations made with the ANU 2.3m telescope 
with HST STIS observations to yield a low resolution spectrum
($R=5000$) over the wide wavelength range $2000 < \lambda
< 8000{\rm
\AA}$.  Savaglio \shortcite{sav98} reported on intermediate-resolution 
($R=11000$) observations obtained with the EMMI spectrograph on the ESO
3.6m telescope in the wavelength region $4500 < \lambda
< 8200{\rm
\AA}$.

In this paper we report on high-resolution observations ($R\sim35000$)
of J2233--606 made with the University College of London Echelle
Spectrograph (UCLES) at the Anglo-Austrlian Telescope (AAT).  These
observations cover the important wavelength region $3530 <
\lambda < 4390{\rm
\AA}$, comprising much of the key Ly$\alpha$\  forest
region.  This data also largely bridges the gap between the
HST STIS
($\lambda < 3500$\AA) observations and the existing EMMI observations
($\lambda >4400$\AA) providing nearly contiguous spectral coverage at
resolution of at least $R=10000$ over the full wavelength range $1150{\rm
\AA} < \lambda < 8200{\rm
\AA}$.

\section{Observations}

\begin{table*}
\begin{minipage}{120mm}
\caption{\rm Details of Observations}
\label{observations}
\begin{tabular}{lccccc}
\hline  
\multicolumn{1}{c}{Date}&Wavelength Range&Grism/CCD&Photometric&Seeing
& Exposure\\
\hline  
1997 August 24&3353--3989\AA&31/Tek&No&2.0--2.5&18000 sec\\
1997 August 25&3353--3989\AA&31/Tek&Yes&1.0--1.2&25200 sec\\
1998 July 31&3319--4075\AA&79/MITLL&No&2.5--4.0&19800 sec\\
1998 August 17&3557--4392\AA&31/Tek&No&1.2--1.4&22200 sec\\
1998 August 18&3557--4392\AA&31/Tek&Yes&1.2--1.5&36000 sec\\
1998 August 19&3557--4392\AA&31/Tek&Yes&1.4--1.5&28800 sec\\
\hline  
\end{tabular}
\end{minipage}
\end{table*}

Observations of J2233--606 were made with UCLES on three separate
observing runs: 1997 August 24--25, 1998 July 28--31 and 1998 August
17--19.  Details are given in table~\ref{observations}.  The
first and third of these runs were both carried out with the 31 line
mm$^{-1}$ cross-dispersing grating and the 1024$^2$ Tektronix CCD.  The 31 line
mm$^{-1}$ grating gives a smaller spacing between the orders
on the echellogram (and thus greater wavelength coverage for a given
detector size), but at the expense of a smaller slit length
(8$\,$arcsec).  The Tektronix chip was binned by a factor of two in
the spatial direction for readout, giving a spatial resolution of 0.7
arcsec$\,$pixel$^{-1}$.  The chip was readout in XTRASLOW mode, with a
readnoise of 2.3 electrons.  Unless interrupted by poor weather, integrations
on the QSO were split into exposures of 3600 seconds, representing a
compromise between short exposures dominated by read noise and long
exposures dominated by cosmic rays events.  The spectral resolution
provided by the Tektronix CCD is $\sim0.04$\AA$\,$pixel$^{-1}$. The
slit-widths used throughout the August 1997 and August 1998
observations (1.2--1.4 arcsec) project to between 2.7 and 3.2 pixels
on this detector.

Good seeing (1.5 arcsec or less) and photometric conditions were
obtained on three out of the five nights with the Tektronix CCD/31
line mm$^{-1}$ grating (see table 1).  This resulted in almost 100000
seconds integration being obtained in good conditions over the key
spectral region $3560 < \lambda < 3990${\AA}, covering
the Ly$\alpha$ emission line and a significant fraction of the
Ly$\alpha$ forest.  In addition, over 70000 seconds integration was
achieved in similar conditions over wavelength region $3990 <
\lambda < 4360$\AA immediately to the red of the Ly$\alpha$
emission line.  Further observations were made in more marginal
conditions.

Unfortunately, we were unable to extend observations to the blue
wavelength limit of the Savaglio observations at $\sim 4400$\AA\  at 
appreciable signal-to-noise ratio (S/N).  The reddest UCLES order lay right on the edge of the 
Tektronix chip ($4360 < \lambda < 4394$\AA), making sky 
subtraction very difficult. We were, however,  able to extract it with 
much higher noise levels.

The 1998 July observations were caried out with the
79 line mm$^{-1}$ grating and a lumogen-coated 4096 $\times$ 2048 
MIT/Lincoln Labs CCD.  Compared to the Tektronix CCD, the
MIT/LL CCD has a significantly lower 
read noise
(1.4 electrons in SLOW mode), is less affected by cosmic rays,
and has a larger area covering more of the echellogram. Unfortunately
much of this additional area is severely vignetting 
(up to 50 per cent) by the existing UCLES camera.
A more serious disadvantage of the MIT/LL CCD
is its much lower quantum efficiency (35 per cent) than the 
Tektronix CCD (40--60 per cent) in the region 3700--5000\AA,
which comprises most of the wavelengths of interest here.
However, it was decided to attempt one set of observations with this
chip to obtain good coverage down in the range 3350--3500\AA, 
where the quantum efficiency of the lumogen-coated MIT/LL CCD 
remains constant at 35 per cent, and is, consequently, much higher 
than that of the Tekronix CCD.  Unfortunately the weather during
the July 1998 run was very poor indeed, only 5 hours observations
were possible, all in poor seeing ($>2.5\,$arcsec).  These 
observations have been added into the final spectrum reproduced
below, with appropriate weights, but provide little additional data.

\section{Data Reduction}

\begin{figure*}
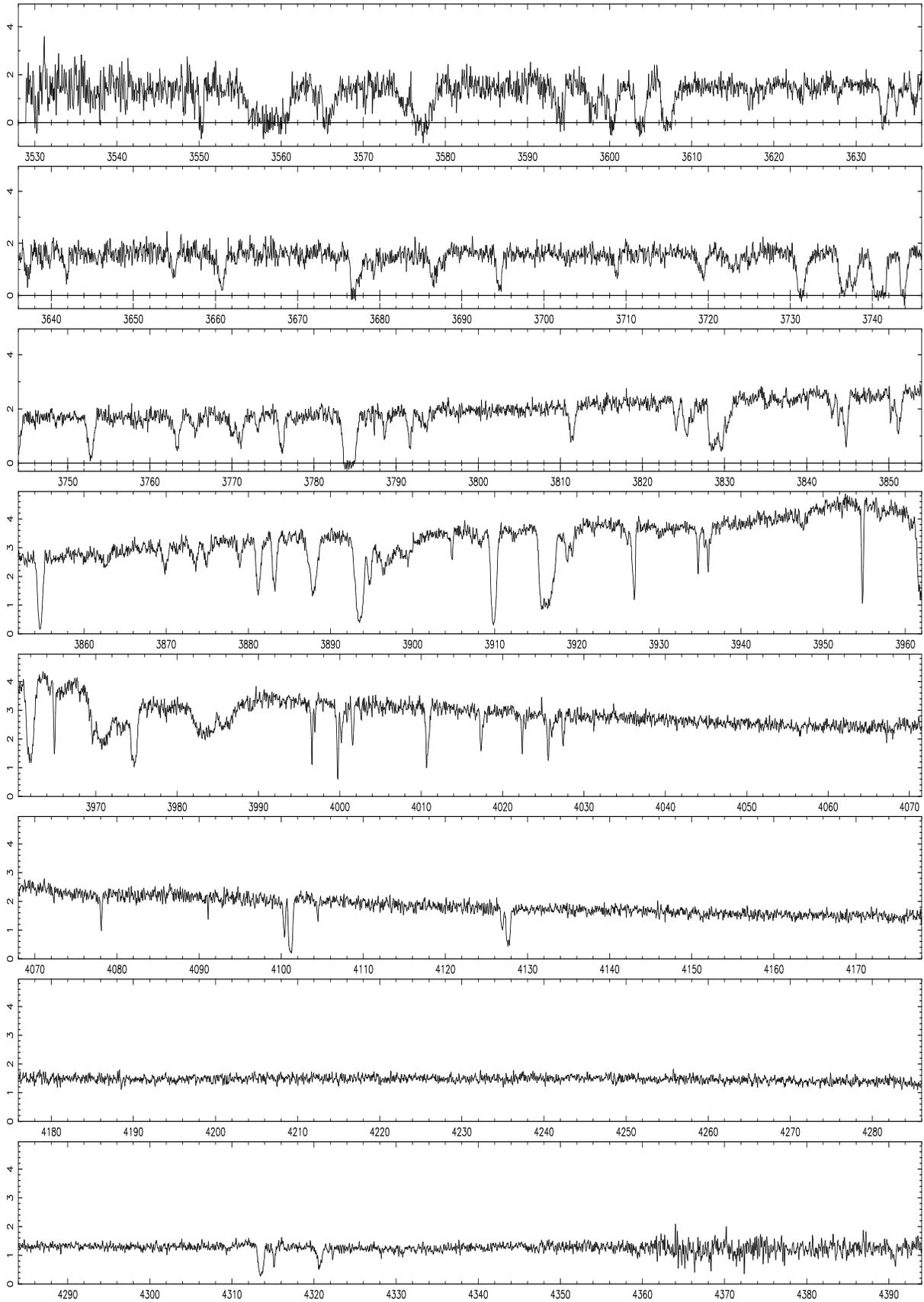

\begin{minipage}{170mm}
\centerline{\hbox{\psfig{figure=plot1.ps,height=28mm,width=160mm}}}
\centerline{\hbox{\psfig{figure=plot2.ps,height=28mm,width=160mm}}}
\centerline{\hbox{\psfig{figure=plot3.ps,height=28mm,width=160mm}}}
\centerline{\hbox{\psfig{figure=plot4.ps,height=28mm,width=160mm}}}
\centerline{\hbox{\psfig{figure=plot5.ps,height=28mm,width=160mm}}}
\centerline{\hbox{\psfig{figure=plot6.ps,height=28mm,width=160mm}}}
\centerline{\hbox{\psfig{figure=plot7.ps,height=28mm,width=160mm}}}
\centerline{\hbox{\psfig{figure=plot8.ps,height=28mm,width=160mm}}}

\caption{The spectrum of J2233-606, plotted against vacuum heliocentric wavelength(\AA).}
\label{figure}
\end{minipage}
\end{figure*}

Standard IRAF packages were used to process the raw images, applying
bias and flat-field corrections. The cosmic rays were flagged using a
median filter and given zero weight in the individual frames. The
sky-subtracted spectra were then optimally extracted, along with a
one-sigma error estimate. Calibration to vacuum heliocentric
wavelengths was obtained using spectra from a Thorium-Argon lamp. The
standard star EG274 was used for flux calibration. The echelle orders
were then resampled to the same dispersion, and added together weighted
according to their S/N. The final spectrum, with a spectral range $3529
< \lambda < 4394$, is shown in figure~\ref{figure}. It has a
resolution, measured from the lamp emission lines, of
$8.5\,$km$\,$s$^{-1}$ full width half maximum (FWHM) and a typical S/N
per $0.05$ bin of $\sim$15 in the Ly$\alpha$\  forest region, and
$\sim$20 long-ward of the Ly$\alpha$\  emission.

Voigt Profiles were fitted, using a $\chi^2$ minimization technique,
to the absorption features in order to determine the redshifts, column
densities and Doppler widths of ions identified with observed
absorption lines. A more complete description of the fitting
procedure, using the software package VPFIT \cite{web87,car92}, can be
found in Outram, Chaffee \& Carswell \shortcite{out98}.

Heavy element lines identified within the Ly$\alpha$\  forest are
either in systems which show features longward of the Ly$\alpha$\
emission line, or were found by looking for common doublets such as
C$\:${\small IV}\ $\lambda\lambda\, 1548, 1550$, Si$\:${\small
IV}\ $\lambda\lambda\, 1393, 1402$, or Mg$\:${\small
II}\ $\lambda\lambda\, 2796, 2803$. 

\begin{table*}
\begin{minipage}{120mm}
\caption{\rm Absorption line parameter list}
\label{list}
\begin{tabular}{lrrrrrrrr}
\hline  
ID & $\lambda_{rest}$ & $\lambda_{obs}$ & $z$ & $\sigma_{z}$ & $b$  & $\sigma_{b}$
 & Log$N$  & $\sigma_{LogN}$\\
\hline  
Si$\:${\small III} & 1206.50 & 3529.85 & 1.925698 & 0.000017 & 14.73 & 2.37 & 12.359 & 0.357\\
Si$\:${\small III} & 1206.50 & 3530.18 & 1.925971 & 0.000003 & 6.75 & 0.37 & 13.300 & 0.483\\
Si$\:${\small III} & 1206.50 & 3549.49 & 1.941975 & 0.000003 & 11.49 & 0.64 & 12.362 & 0.424\\
Si$\:${\small III} & 1206.50 & 3550.05 & 1.942434 & 0.000005 & 10.71 & 0.67 & 13.039 & 0.258\\
Si$\:${\small III} & 1206.50 & 3550.26 & 1.942610 & 0.000002 & 6.42 & 0.34 & 14.988 & 0.478\\
H$\:${\small I} & 1215.67 & 3556.33 & 1.925409 & 0.000054 & 53.28 & 7.04 & 14.148 & 0.071\\
H$\:${\small I} & 1215.67 & 3558.16 & 1.926913 & 0.000027 & 30.44 & 3.99 & 17.780 & 0.845\\
H$\:${\small I} & 1215.67 & 3560.23 & 1.928615 & 0.000027 & 40.78 & 5.68 & 14.887 & 0.246\\
H$\:${\small I} & 1215.67 & 3565.69 & 1.933103 & 0.000027 & 59.33 & 3.70 & 14.276 & 0.042\\
H$\:${\small I} & 1215.67 & 3570.20 & 1.936817 & 0.000034 & 20.03 & 4.81 & 13.194 & 0.095\\
H$\:${\small I} & 1215.67 & 3574.59 & 1.940430 & 0.000187 & 43.99 & 18.38 & 13.498 & 0.233\\
H$\:${\small I} & 1215.67 & 3575.10 & 1.940845 & 0.000028 & 16.47 & 5.49 & 13.493 & 0.200\\
H$\:${\small I} & 1215.67 & 3577.02 & 1.942429 & 0.000018 & 23.44 & 0.92 & 18.205 & 0.099\\
H$\:${\small I} & 1215.67 & 3578.45 & 1.943607 & 0.000036 & 19.38 & 5.33 & 13.360 & 0.119\\
H$\:${\small I} & 1215.67 & 3581.13 & 1.945809 & 0.000103 & 44.23 & 15.49 & 13.128 & 0.142\\
H$\:${\small I} & 1215.67 & 3593.20 & 1.955738 & 0.000440 & 38.93 & 38.92 & 13.137 & 0.633\\
H$\:${\small I} & 1215.67 & 3593.98 & 1.956380 & 0.000064 & 33.97 & 6.00 & 14.007 & 0.100\\
H$\:${\small I} & 1215.67 & 3597.95 & 1.959642 & 0.000033 & 56.65 & 4.62 & 13.930 & 0.038\\
H$\:${\small I} & 1215.67 & 3600.19 & 1.961487 & 0.000021 & 46.16 & 2.99 & 14.197 & 0.045\\
H$\:${\small I} & 1215.67 & 3603.71 & 1.964382 & 0.000014 & 33.42 & 3.07 & 14.820 & 0.195\\
H$\:${\small I} & 1215.67 & 3607.05 & 1.967126 & 0.000016 & 42.40 & 3.10 & 14.641 & 0.115\\
H$\:${\small I} & 1215.67 & 3616.94 & 1.975263 & 0.000012 & 10.62 & 1.66 & 13.285 & 0.071\\
H$\:${\small I} & 1215.67 & 3623.22 & 1.980434 & 0.000042 & 29.53 & 6.16 & 13.094 & 0.081\\
H$\:${\small I} & 1215.67 & 3627.82 & 1.984215 & 0.000035 & 28.02 & 4.95 & 13.175 & 0.069\\
H$\:${\small I} & 1215.67 & 3633.34 & 1.988755 & 0.000011 & 28.17 & 1.78 & 14.175 & 0.065\\
H$\:${\small I} & 1215.67 & 3634.94 & 1.990068 & 0.000014 & 20.30 & 1.88 & 13.529 & 0.042\\
H$\:${\small I} & 1215.67 & 3637.11 & 1.991857 & 0.000019 & 25.34 & 2.53 & 13.511 & 0.043\\
H$\:${\small I} & 1215.67 & 3641.82 & 1.995729 & 0.000015 & 18.05 & 2.03 & 13.322 & 0.046\\
H$\:${\small I} & 1215.67 & 3654.90 & 2.006494 & 0.000022 & 28.39 & 2.93 & 13.462 & 0.043\\
H$\:${\small I} & 1215.67 & 3659.74 & 2.010472 & 0.000064 & 32.16 & 9.31 & 13.100 & 0.107\\
H$\:${\small I} & 1215.67 & 3660.74 & 2.011293 & 0.000018 & 31.05 & 2.47 & 13.860 & 0.036\\
H$\:${\small I} & 1215.67 & 3670.80 & 2.019568 & 0.000019 & 9.37 & 2.80 & 12.741 & 0.102\\
H$\:${\small I} & 1215.67 & 3676.76 & 2.024476 & 0.000018 & 19.30 & 2.40 & 14.217 & 0.143\\
H$\:${\small I} & 1215.67 & 3677.39 & 2.024992 & 0.000097 & 41.32 & 8.95 & 13.689 & 0.114\\
H$\:${\small I} & 1215.67 & 3678.59 & 2.025979 & 0.000023 & 10.74 & 3.53 & 12.717 & 0.112\\
H$\:${\small I} & 1215.67 & 3679.21 & 2.026488 & 0.000019 & 20.07 & 2.73 & 13.236 & 0.052\\
H$\:${\small I} & 1215.67 & 3686.52 & 2.032498 & 0.000024 & 41.26 & 3.32 & 13.733 & 0.031\\
H$\:${\small I} & 1215.67 & 3687.61 & 2.033399 & 0.000048 & 25.94 & 6.88 & 12.971 & 0.101\\
H$\:${\small I} & 1215.67 & 3694.53 & 2.039087 & 0.000010 & 24.44 & 1.31 & 13.826 & 0.032\\
Si$\:${\small II} & 1260.42 & 3708.71 & 1.942434 & 0.000005 & 10.71 & 0.67 & 12.577 & 0.067\\
Si$\:${\small II} & 1260.42 & 3708.93 & 1.942610 & 0.000002 & 6.42 & 0.34 & 12.524 & 0.073\\
H$\:${\small I} & 1215.67 & 3731.28 & 2.069322 & 0.000012 & 28.55 & 2.34 & 14.234 & 0.069\\
H$\:${\small I} & 1215.67 & 3731.57 & 2.069556 & 0.000067 & 85.86 & 9.48 & 13.694 & 0.068\\
H$\:${\small I} & 1215.67 & 3736.45 & 2.073570 & 0.000013 & 35.52 & 1.72 & 14.140 & 0.028\\
H$\:${\small I} & 1215.67 & 3737.66 & 2.074572 & 0.000018 & 39.47 & 2.71 & 13.880 & 0.027\\
H$\:${\small I} & 1215.67 & 3740.23 & 2.076684 & 0.000923 & 109.04 & 56.41 & 13.559 & 0.496\\
H$\:${\small I} & 1215.67 & 3740.83 & 2.077173 & 0.000011 & 33.08 & 3.46 & 15.400 & 0.255\\
H$\:${\small I} & 1215.67 & 3742.36 & 2.078432 & 0.000272 & 37.21 & 29.70 & 12.785 & 0.672\\
H$\:${\small I} & 1215.67 & 3743.75 & 2.079577 & 0.000007 & 24.78 & 1.19 & 14.211 & 0.051\\
H$\:${\small I} & 1215.67 & 3752.79 & 2.087013 & 0.000008 & 30.05 & 1.07 & 13.894 & 0.021\\
H$\:${\small I} & 1215.67 & 3758.69 & 2.091871 & 0.000074 & 75.92 & 11.84 & 13.233 & 0.073\\
H$\:${\small I} & 1215.67 & 3763.27 & 2.095634 & 0.000011 & 29.43 & 1.34 & 13.613 & 0.020\\
H$\:${\small I} & 1215.67 & 3765.50 & 2.097465 & 0.000022 & 31.02 & 2.96 & 13.202 & 0.037\\
H$\:${\small I} & 1215.67 & 3769.99 & 2.101162 & 0.000022 & 26.94 & 2.99 & 13.273 & 0.042\\
H$\:${\small I} & 1215.67 & 3770.95 & 2.101952 & 0.000017 & 34.09 & 2.38 & 13.584 & 0.025\\
H$\:${\small I} & 1215.67 & 3773.12 & 2.103734 & 0.000017 & 19.84 & 2.25 & 13.053 & 0.043\\
H$\:${\small I} & 1215.67 & 3776.04 & 2.106139 & 0.000009 & 26.24 & 1.09 & 13.688 & 0.020\\
H$\:${\small I} & 1215.67 & 3784.29 & 2.112923 & 0.000007 & 43.05 & 1.35 & 14.800 & 0.059\\
H$\:${\small I} & 1215.67 & 3788.59 & 2.116465 & 0.000011 & 17.33 & 1.40 & 13.192 & 0.032\\
H$\:${\small I} & 1215.67 & 3791.63 & 2.118963 & 0.000009 & 22.18 & 1.14 & 13.476 & 0.022\\
H$\:${\small I} & 1215.67 & 3793.45 & 2.120458 & 0.000036 & 53.52 & 4.73 & 13.313 & 0.034\\
H$\:${\small I} & 1215.67 & 3811.37 & 2.135201 & 0.000009 & 30.67 & 1.17 & 13.550 & 0.016\\
H$\:${\small I} & 1215.67 & 3824.09 & 2.145662 & 0.000009 & 20.59 & 1.20 & 13.208 & 0.022\\
\hline  
\end{tabular}
\end{minipage}
\end{table*}
\setcounter{table}{1}
\begin{table*}
\begin{minipage}{120mm}
\caption{\rm Absorption line parameter list}
\begin{tabular}{lrrrrrrrr}
\hline  
ID & $\lambda_{rest}$ & $\lambda_{obs}$ & $z$ & $\sigma_{z}$ & $b$  & $\sigma_{b}$
 & Log$N$  & $\sigma_{LogN}$\\
\hline  
H$\:${\small I} & 1215.67 & 3825.34 & 2.146696 & 0.000010 & 30.52 & 1.40 & 13.523 & 0.016\\
H$\:${\small I} & 1215.67 & 3826.10 & 2.147319 & 0.000012 & 15.67 & 1.78 & 12.965 & 0.043\\
H$\:${\small I} & 1215.67 & 3826.96 & 2.148028 & 0.000034 & 32.62 & 5.21 & 12.866 & 0.055\\
H$\:${\small I} & 1215.67 & 3828.30 & 2.149128 & 0.000071 & 24.79 & 6.66 & 13.591 & 0.492\\
H$\:${\small I} & 1215.67 & 3828.86 & 2.149590 & 0.000191 & 39.99 & 28.86 & 13.803 & 0.370\\
H$\:${\small I} & 1215.67 & 3829.58 & 2.150181 & 0.000017 & 18.35 & 2.53 & 13.588 & 0.124\\
H$\:${\small I} & 1215.67 & 3830.27 & 2.150747 & 0.000038 & 44.44 & 4.12 & 13.517 & 0.040\\
H$\:${\small I} & 1215.67 & 3835.20 & 2.154801 & 0.000029 & 19.23 & 4.18 & 12.613 & 0.083\\
Si$\:${\small II} & 1304.37 & 3838.02 & 1.942434 & 0.000005 & 10.71 & 0.67 & 12.577 & 0.067\\
Si$\:${\small II} & 1304.37 & 3838.25 & 1.942610 & 0.000002 & 6.42 & 0.34 & 12.524 & 0.073\\
H$\:${\small I} & 1215.67 & 3843.06 & 2.161271 & 0.000015 & 25.23 & 2.12 & 13.041 & 0.032\\
C$\:${\small IV} & 1548.20 & 3843.81 & 1.482769 & 0.000003 & 5.16 & 0.83 & 13.087 & 0.069\\
C$\:${\small IV} & 1548.20 & 3844.08 & 1.482942 & 0.000019 & 18.61 & 4.30 & 13.167 & 0.084\\
C$\:${\small IV} & 1548.20 & 3844.75 & 1.483374 & 0.000004 & 21.58 & 0.76 & 13.817 & 0.013\\
C$\:${\small IV} & 1550.77 & 3850.20 & 1.482769 & 0.000003 & 5.16 & 0.83 & 13.087 & 0.069\\
C$\:${\small IV} & 1550.77 & 3850.47 & 1.482942 & 0.000019 & 18.61 & 4.30 & 13.167 & 0.084\\
C$\:${\small IV} & 1550.77 & 3851.14 & 1.483374 & 0.000004 & 21.58 & 0.76 & 13.817 & 0.013\\
H$\:${\small I} & 1215.67 & 3854.62 & 2.170781 & 0.000004 & 21.92 & 0.52 & 13.861 & 0.015\\
H$\:${\small I} & 1215.67 & 3862.53 & 2.177281 & 0.000027 & 42.92 & 3.52 & 13.025 & 0.031\\
H$\:${\small I} & 1215.67 & 3869.87 & 2.183320 & 0.000012 & 28.99 & 1.54 & 13.111 & 0.020\\
H$\:${\small I} & 1215.67 & 3873.43 & 2.186253 & 0.000017 & 38.66 & 2.26 & 13.143 & 0.022\\
H$\:${\small I} & 1215.67 & 3874.89 & 2.187449 & 0.000014 & 22.33 & 1.89 & 12.849 & 0.031\\
H$\:${\small I} & 1215.67 & 3878.93 & 2.190779 & 0.000012 & 25.59 & 1.58 & 12.988 & 0.023\\
H$\:${\small I} & 1215.67 & 3881.21 & 2.192647 & 0.000005 & 24.94 & 0.62 & 13.453 & 0.010\\
H$\:${\small I} & 1215.67 & 3883.18 & 2.194270 & 0.000005 & 22.14 & 0.64 & 13.350 & 0.012\\
H$\:${\small I} & 1215.67 & 3887.91 & 2.198162 & 0.000003 & 43.99 & 1.01 & 13.738 & 0.011\\
H$\:${\small I} & 1215.67 & 3889.30 & 2.199303 & 0.000056 & 42.93 & 11.88 & 12.716 & 0.207\\
H$\:${\small I} & 1215.67 & 3891.44 & 2.201069 & 0.000048 & 159.01 & 50.56 & 13.389 & 0.118\\
H$\:${\small I} & 1215.67 & 3893.52 & 2.202775 & 0.000004 & 41.08 & 0.75 & 14.064 & 0.011\\
H$\:${\small I} & 1215.67 & 3894.74 & 2.203777 & 0.000007 & 22.66 & 1.05 & 13.318 & 0.023\\
H$\:${\small I} & 1215.67 & 3896.55 & 2.205269 & 0.000064 & 108.28 & 8.44 & 13.826 & 0.041\\
H$\:${\small I} & 1215.67 & 3898.99 & 2.207278 & 0.000037 & 64.71 & 22.18 & 13.133 & 0.525\\
H$\:${\small I} & 1215.67 & 3899.56 & 2.207746 & 0.000494 & 115.62 & 17.23 & 13.489 & 0.297\\
C$\:${\small II} & 1334.53 & 3904.44 & 1.925698 & 0.000017 & 22.53 & 2.37 & 12.788 & 0.081\\
C$\:${\small II} & 1334.53 & 3904.80 & 1.925971 & 0.000003 & 10.32 & 0.37 & 13.124 & 0.031\\
H$\:${\small I} & 1215.67 & 3908.28 & 2.214920 & 0.000027 & 27.31 & 3.60 & 12.614 & 0.048\\
H$\:${\small I} & 1215.67 & 3909.86 & 2.216218 & 0.000003 & 25.14 & 0.35 & 13.905 & 0.009\\
H$\:${\small I} & 1215.67 & 3915.65 & 2.220982 & 0.000008 & 19.71 & 1.35 & 13.372 & 0.041\\
H$\:${\small I} & 1215.67 & 3916.44 & 2.221628 & 0.000013 & 63.12 & 1.21 & 14.080 & 0.011\\
H$\:${\small I} & 1215.67 & 3918.87 & 2.223629 & 0.000022 & 44.80 & 2.80 & 13.266 & 0.026\\
H$\:${\small I} & 1215.67 & 3919.46 & 2.224118 & 0.000009 & 10.04 & 1.75 & 12.536 & 0.082\\
C$\:${\small II} & 1334.53 & 3926.16 & 1.941975 & 0.000003 & 17.57 & 0.64 & 13.165 & 0.025\\
C$\:${\small II} & 1334.53 & 3926.77 & 1.942434 & 0.000005 & 16.38 & 0.67 & 13.482 & 0.028\\
C$\:${\small II} & 1334.53 & 3927.01 & 1.942610 & 0.000002 & 9.82 & 0.34 & 13.631 & 0.019\\
Ca$\:${\small II} & 3934.78 & 3934.77 & -0.000003 & 0.000001 & 8.42 & 0.41 & 12.223 & 0.016\\
H$\:${\small I} & 1215.67 & 3935.92 & 2.237659 & 0.000011 & 31.74 & 1.40 & 13.147 & 0.016\\
H$\:${\small I} & 1215.67 & 3947.50 & 2.247180 & 0.000030 & 46.97 & 3.97 & 12.890 & 0.031\\
Mg$\:${\small II} & 2796.35 & 3954.77 & 0.414261 & 0.000001 & 7.31 & 0.18 & 12.809 & 0.011\\
N$\:${\small V} & 1238.82 & 3960.71 & 2.197161 & 0.000026 & 32.51 & 4.06 & 13.122 & 0.048\\
N$\:${\small V} & 1238.82 & 3961.95 & 2.198162 & 0.000003 & 31.46 & 0.58 & 14.568 & 0.026\\
Mg$\:${\small II} & 2803.53 & 3964.92 & 0.414261 & 0.000001 & 7.31 & 0.18 & 12.809 & 0.011\\
N$\:${\small V} & 1238.82 & 3965.55 & 2.201069 & 0.000048 & 96.15 & 7.61 & 13.580 & 0.041\\
Ca$\:${\small II} & 3969.59 & 3969.58 & -0.000003 & 0.000001 & 8.42 & 0.41 & 12.223 & 0.016\\
N$\:${\small V} & 1238.82 & 3970.75 & 2.205269 & 0.000064 & 101.12 & 6.88 & 14.361 & 0.094\\
N$\:${\small V} & 1238.82 & 3973.24 & 2.207278 & 0.000037 & 58.75 & 7.26 & 13.779 & 0.135\\
N$\:${\small V} & 1242.80 & 3973.43 & 2.197161 & 0.000026 & 32.51 & 4.06 & 13.122 & 0.048\\
N$\:${\small V} & 1238.82 & 3973.82 & 2.207746 & 0.000494 & 193.86 & 129.48 & 13.992 & 0.596\\
N$\:${\small V} & 1242.80 & 3974.68 & 2.198162 & 0.000003 & 31.46 & 0.58 & 14.568 & 0.026\\
N$\:${\small V} & 1238.82 & 3976.65 & 2.210034 & 0.001863 & 221.92 & 201.82 & 13.708 & 0.847\\
N$\:${\small V} & 1242.80 & 3978.29 & 2.201069 & 0.000048 & 96.15 & 7.61 & 13.580 & 0.041\\
N$\:${\small V} & 1242.80 & 3983.51 & 2.205269 & 0.000064 & 101.12 & 6.88 & 14.361 & 0.094\\
N$\:${\small V} & 1242.80 & 3986.01 & 2.207278 & 0.000037 & 58.75 & 7.26 & 13.779 & 0.135\\
N$\:${\small V} & 1242.80 & 3986.58 & 2.207746 & 0.000494 & 193.86 & 129.48 & 13.992 & 0.596\\
\hline  
\end{tabular}
\end{minipage}
\end{table*}
\setcounter{table}{1}
\begin{table*}
\begin{minipage}{120mm}
\caption{\rm Absorption line parameter list}
\begin{tabular}{lrrrrrrrr}
\hline  
ID & $\lambda_{rest}$ & $\lambda_{obs}$ & $z$ & $\sigma_{z}$ & $b$  & $\sigma_{b}$
 & Log$N$  & $\sigma_{LogN}$\\
\hline  
N$\:${\small V} & 1242.80 & 3989.43 & 2.210034 & 0.001863 & 221.92 & 201.82 & 13.708 & 0.847\\
Si$\:${\small IV} & 1393.76 & 3996.57 & 1.867485 & 0.000002 & 6.07 & 0.27 & 12.905 & 0.015\\
Si$\:${\small IV} & 1393.76 & 3996.93 & 1.867745 & 0.000004 & 5.38 & 0.67 & 12.373 & 0.033\\
Si$\:${\small IV} & 1393.76 & 3999.52 & 1.869598 & 0.000040 & 10.39 & 4.48 & 12.167 & 0.206\\
Si$\:${\small IV} & 1393.76 & 3999.75 & 1.869762 & 0.000003 & 7.07 & 0.53 & 13.122 & 0.028\\
Si$\:${\small IV} & 1393.76 & 4000.15 & 1.870055 & 0.000008 & 8.46 & 2.22 & 12.523 & 0.183\\
Si$\:${\small IV} & 1393.76 & 4000.32 & 1.870175 & 0.000066 & 23.00 & 5.09 & 12.643 & 0.174\\
Si$\:${\small IV} & 1393.76 & 4000.88 & 1.870576 & 0.000006 & 5.03 & 1.20 & 12.118 & 0.062\\
Si$\:${\small IV} & 1393.76 & 4001.58 & 1.871076 & 0.000003 & 9.97 & 0.48 & 12.833 & 0.017\\
Si$\:${\small IV} & 1393.76 & 4002.60 & 1.871813 & 0.000010 & 8.08 & 1.73 & 12.112 & 0.070\\
C$\:${\small IV} & 1548.20 & 4010.66 & 1.590541 & 0.000003 & 4.62 & 0.82 & 13.002 & 0.060\\
C$\:${\small IV} & 1548.20 & 4010.79 & 1.590621 & 0.000007 & 18.82 & 0.80 & 13.440 & 0.026\\
C$\:${\small IV} & 1550.77 & 4017.33 & 1.590541 & 0.000003 & 4.62 & 0.82 & 13.002 & 0.060\\
C$\:${\small IV} & 1550.77 & 4017.46 & 1.590621 & 0.000007 & 18.82 & 0.80 & 13.440 & 0.026\\
Si$\:${\small IV} & 1402.77 & 4022.42 & 1.867485 & 0.000002 & 6.07 & 0.27 & 12.905 & 0.015\\
Si$\:${\small IV} & 1402.77 & 4022.79 & 1.867745 & 0.000004 & 5.38 & 0.67 & 12.373 & 0.033\\
Si$\:${\small IV} & 1402.77 & 4025.39 & 1.869598 & 0.000040 & 10.39 & 4.48 & 12.167 & 0.206\\
Si$\:${\small IV} & 1402.77 & 4025.62 & 1.869762 & 0.000003 & 7.07 & 0.53 & 13.122 & 0.028\\
Si$\:${\small IV} & 1402.77 & 4026.03 & 1.870055 & 0.000008 & 8.46 & 2.22 & 12.523 & 0.183\\
Si$\:${\small IV} & 1402.77 & 4026.20 & 1.870175 & 0.000066 & 23.00 & 5.09 & 12.643 & 0.174\\
Si$\:${\small IV} & 1402.77 & 4026.76 & 1.870576 & 0.000006 & 5.03 & 1.20 & 12.118 & 0.062\\
Si$\:${\small IV} & 1402.77 & 4027.46 & 1.871076 & 0.000003 & 9.97 & 0.48 & 12.833 & 0.017\\
Si$\:${\small IV} & 1402.77 & 4028.49 & 1.871813 & 0.000010 & 8.08 & 1.73 & 12.112 & 0.070\\
Si$\:${\small IV} & 1393.76 & 4077.71 & 1.925698 & 0.000017 & 14.73 & 2.37 & 12.318 & 0.059\\
Si$\:${\small IV} & 1393.76 & 4078.09 & 1.925971 & 0.000003 & 6.75 & 0.37 & 12.775 & 0.021\\
Si$\:${\small IV} & 1393.76 & 4100.39 & 1.941975 & 0.000003 & 11.49 & 0.64 & 13.040 & 0.016\\
Si$\:${\small IV} & 1393.76 & 4101.03 & 1.942434 & 0.000005 & 10.71 & 0.67 & 13.426 & 0.023\\
Si$\:${\small IV} & 1393.76 & 4101.28 & 1.942610 & 0.000002 & 6.42 & 0.34 & 13.341 & 0.030\\
Si$\:${\small IV} & 1402.77 & 4104.08 & 1.925698 & 0.000017 & 14.73 & 2.37 & 12.318 & 0.059\\
Si$\:${\small IV} & 1402.77 & 4104.46 & 1.925971 & 0.000003 & 6.75 & 0.37 & 12.775 & 0.021\\
Si$\:${\small IV} & 1402.77 & 4126.91 & 1.941975 & 0.000003 & 11.49 & 0.64 & 13.040 & 0.016\\
Si$\:${\small IV} & 1402.77 & 4127.56 & 1.942434 & 0.000005 & 10.71 & 0.67 & 13.426 & 0.023\\
Si$\:${\small IV} & 1402.77 & 4127.81 & 1.942610 & 0.000002 & 6.42 & 0.34 & 13.341 & 0.030\\
C$\:${\small IV} & 1548.20 & 4309.48 & 1.783553 & 0.000024 & 29.66 & 4.14 & 13.051 & 0.060\\
C$\:${\small IV} & 1548.20 & 4313.52 & 1.786161 & 0.000004 & 23.43 & 0.57 & 13.915 & 0.011\\
C$\:${\small IV} & 1548.20 & 4314.57 & 1.786841 & 0.000009 & 6.12 & 1.69 & 12.572 & 0.077\\
C$\:${\small IV} & 1548.20 & 4315.13 & 1.787201 & 0.000004 & 8.03 & 0.65 & 13.169 & 0.027\\
C$\:${\small IV} & 1550.77 & 4316.65 & 1.783553 & 0.000024 & 29.66 & 4.14 & 13.051 & 0.060\\
C$\:${\small IV} & 1550.77 & 4320.69 & 1.786161 & 0.000004 & 23.43 & 0.57 & 13.915 & 0.011\\
C$\:${\small IV} & 1550.77 & 4321.75 & 1.786841 & 0.000009 & 6.12 & 1.69 & 12.572 & 0.077\\
C$\:${\small IV} & 1550.77 & 4322.31 & 1.787201 & 0.000004 & 8.03 & 0.65 & 13.169 & 0.027\\
?? & 1215.67 & 4390.66 & 2.611718 & 0.000036 & 20.77 & 4.32 & 13.063 & 0.082\\
\hline  
\end{tabular}
\end{minipage}
\end{table*}

\section{Results}
We present the absorption line parameter fits in table~\ref{list}.  A
total of 10 heavy element systems, as well as 89 Ly$\alpha$\  forest
lines were seen in the spectrum. The HI column density limit in the
forest is set mainly by continuum uncertainties, and varies with S/N.
On average the limit is log $N($H I$)\sim 13$ for a Doppler parameter
of 20 km$\,$s$^{-1}$, corresponding to an equivalent width limit of
$\sim25$m\AA.

\subsection{The Ly$\alpha$\  Forest}

We have obtained spectral coverage of the Ly$\alpha$\  forest over the wavelength region 3530\ $ < \lambda <$ 3950\AA\ , corresponding to a redshifts from $z=1.9$ to the emission redshift of the quasar, $z=2.24$. The S/N per 0.05\AA\  bin falls off at shorter wavelengths, dropping gradually from $\sim$30 on the Ly$\alpha$\  emission peak, to approximately six at $\lambda =$ 3600\AA. Below $\lambda =$ 3550\AA\ the S/N drops dramatically to around three, because this region was not covered in the August 1998 run. It was included, however, because of the important Si$\:${\small III} lines corresponding to the $z=1.92$ and $z=1.94$ systems. No other Ly$\alpha$\  lines were detected here. 

A total of 89 Ly$\alpha$\  lines were fitted. The parameter distributions follow a similar pattern to that seen in other quasar spectra (e.g. Lu et al. 1996; Kirkman \& Tytler 1997). The H I column density distribution, for 13.3 $<$ log $N <$ 16, is consistent with a power law distribution of slope $-1.5$. The Doppler parameter distribution is well described by a Gaussian of mean $b=25\,$km$\,$s$^{-1}$, $\sigma_{b} =$15 km$\,$s$^{-1}$, and a cut-off below  $b=18\,$km$\,$s$^{-1}$. About 8 per cent of the Ly$\alpha$\  lines have Doppler widths smaller than this, but this could be due to either contamination of the sample by narrower unidentified heavy element lines, or simply noise features. Indeed all of these lines have column densities log $N <$ 13.5, close to the local detection limit. There is no observed correlation between the Doppler widths and column densities of the lines.

\subsection{The Intervening Heavy Element Systems}
A total of 78 heavy element lines have been fitted from eight
intervening, and two associated heavy element systems. One further line
was observed at 4390.66\AA, right at the red end of the spectrum. It is
not associated with any known heavy element system, and is possibly
either C$\:${\small IV} $\lambda\, 1548 $ at redshift $z=1.836$, or
Mg$\:${\small II} $\lambda\, 2796  $ at redshift $z=0.570$, although
other identifications are possible. The profile fit was performed with
the assumption that the line has the rest wavelength and oscillator
strength of Ly$\alpha$.

\subsubsection{ $z=0.0000$}
 Narrow Ca$\:${\small II} $\lambda\lambda\, 3934, 3969  $ is seen from the interstellar medium in the Galaxy.

\subsubsection{ $z=0.4143$}
 A single Mg$\:${\small II} $\lambda\lambda\, 2796, 2803  $ doublet occurs at this redshift, with Doppler parameter $b=7\,$km$\,$s$^{-1}$ and column density log $N$(Mg$\:${\small II}) = 13.1.

\subsubsection{ $z=1.483$}
A complex of three C$\:${\small IV} $\lambda\lambda\, 1548, 1550  $ doublets, spanning 75 km$\,$s$^{-1}$, is seen within the Ly$\alpha$\  forest. Although the C$\:${\small IV} complex is observed in the Ly$\alpha$\  forest, and has no further confirming heavy element lines, the system has a distinctive pattern that makes this a firm identification.

\subsubsection{ Not $z=1.5034$}
Savaglio \shortcite{sav98} reported the tentative identification of a
Mg$\:${\small II} doublet at $z=1.5034$. However, no corresponding
Ly$\alpha$\ absorption line is seen in the STIS data and no
C$\:${\small IV} absorption is visible in our spectrum, making the
original identification very insecure.

\subsubsection{ $z=1.591$}
A C$\:${\small IV} $\lambda\lambda\, 1548, 1550  $ feature, consisting 2 lines separated by $9\,$km$\,$s$^{-1}$ was seen at 4011 and  4018\AA, redward of the Ly$\alpha$\  emission line. No Si$\:${\small IV} absorption was detected for this system;  log $N($Si$\:${\small IV}$) < 12.3$ is a $4 \sigma$ upper limit to the column density.

\subsubsection{ $z=1.786$}
Sealey et al. \shortcite{sdw98} first reported C$\:${\small IV} associated with a prominent Ly$\alpha$\  line in the STIS data. Savaglio \shortcite{sav98} detected the C$\:${\small IV} absorption at high resolution, but very low S/N. MgI $\lambda\lambda\, 2852  $ was also detected. The UCLES spectrum shows four C$\:${\small IV} $\lambda\lambda\, 1548, 1550  $ components spanning almost 400 km$\,$s$^{-1}$. Any Si$\:${\small IV}  
$\lambda\lambda\, 1393, 1402  $ absorption for this system, if significant, was blended with Ly$\alpha$\  lines in the forest, making it impossible to determine a reliable column density.

\subsubsection{ $z=1.87$}
This heavy element system has a complex structure of nine Si$\:${\small IV} $ \lambda\lambda\, 1393, 1402  $ components, spanning 450 km$\,$s$^{-1}$. Savaglio \shortcite{sav98} detected complex C$\:${\small IV} absorption for this system, but noted that much higher S/N would be required to deblend the lines. She also detected MgI $ \lambda\, 2853  $, but the identification of this line (normally only seen in damped Ly$\alpha$\  systems) is doubtful.

\subsubsection{ $z=1.925$}
The UCLES spectrum shows a large saturated Ly$\alpha$\  line, fitted by three components, at 3560\AA. Without the constraint of higher order Lyman series absorption lines, the column density for this system is very poorly constrained. C$\:${\small II} $\lambda\, 1334  $, Si$\:${\small III} $\lambda\, 1206 $, and Si$\:${\small IV} $\lambda\lambda\, 1393, 1402  $ absorption lines were detected for this system. 2 heavy element components were seen, at $z=1.9256$ and $z=1.9259$. This is over 100 km$\,$s$^{-1}$ blueward of the centre of the saturated Ly$\alpha$\  line; however no heavy element absorption was detected for the central hydrogen component. 

Sealey et al. \shortcite{sdw98} investigated this system as a possible cause of the Lyman break seen in the STIS spectrum at $\lambda\, \sim 2700  $, concluding however that the system at $z=1.942$ is more likely to be the major contributor. Savaglio \shortcite{sav98} searched for heavy element lines associated with this system in the EMMI spectrum, finding no associated C$\:${\small IV} absorption, and only an upper limit to the column density of log $N($C$\:${\small IV}$) < 13.7$. This implies that the column density ratios Si$\:${\small IV} / C$\:${\small IV} $\ga 0.15$, and C$\:${\small II} / C$\:${\small IV} $\ga 0.4$, both of which are relatively high \cite{son98}. Savaglio tentatively identified FeI$\lambda\, 2484  $ and Fe$\:${\small II}$\lambda\, 2382  $ in the EMMI spectrum. Further investigation of the spectrum at the exact redshift of the observed heavy element lines shows that these identifications are doubtful.

\subsubsection{ $z=1.942$}
The Ly$\alpha$\  line is present in the UCLES spectrum (at 3577\AA), but it is saturated, and too noisy to determine an accurate column density directly. If solely responsible for the Lyman break seen in the STIS spectrum, however, this system has a column density of log $N($H$\:${\small I}$) \sim 17.5$. C$\:${\small II} $\lambda\, 1334  $, Si$\:${\small II} $\lambda\lambda\, 1260, 1304  $, Si$\:${\small III} $\lambda\, 1206  $, and Si$\:${\small IV} $\lambda\lambda\, 1393, 1402  $ were all detected in the spectrum, and three components were fitted, at $z=1.94197$, $z=1.94243$, and $z=1.94261$. No OI $\lambda\, 1302  $ absorption was detected; a $4 \sigma$ upper limit to the OI column density is log $N($O I$) < 12.95$.

Sealey et al. suggested that this system may be a good candidate for a measurement of the D / H ratio. Unfortunately, the weak $z=1.94197$ line lies 60 km$\,$s$^{-1}$ blueward of the other two stronger lines, and the hydrogen cloud associated with this line will probably adversely affect any determination of the deuterium abundance from the main system. 

Savaglio \shortcite{sav98} identified strong C$\:${\small IV} $\lambda\lambda\, 1548, 1550  $ absorption at the same redshift as the absorption features in the UCLES data. Mg$\:${\small II} $\lambda\lambda\, 2796, 2803  $, and Al$\:${\small III} $\lambda\lambda\, 1854, 1862  $ were also detected for the strongest component.

\subsubsection{ $z=2.077$}

This heavy element system, identified by Savaglio \shortcite{sav98}, has a  weak C$\:${\small IV}$\lambda\lambda\, 1548, 1550  $ doublet. The corresponding Ly$\alpha$\  line is seen at $z=2.07717$ and has a log column density of 15.4. No associated heavy element lines were found in the UCLES spectrum, and a $4 \sigma$ upper limit to the Si$\:${\small IV} column density is log $N($Si$\:${\small IV}$) < 11.9$. 

\subsection{The Associated Heavy Element Systems}
The UCLES spectrum shows the clear signature of strong, broad N$\:${\small V} absorption in the region 3960\ $ < \lambda <$ 3990\AA. The corresponding H I absorption, at 3885\ $ < \lambda <$ 3900\AA, is also broad, and apparently not saturated, indicative of highly ionized systems.

\subsubsection{ $z=2.198$}

A strong N$\:${\small V} $\lambda\lambda\, 1238, 1242  $ doublet, and corresponding Ly$\alpha$\  line are seen at $z=2.198$. The N$\:${\small V} doublet ratio is roughly 1:1, yet the lines have residual intensity $\sim 35$ per cent of the continuum. The likely explanation for this is that there is incomplete coverage of the background source, as has been seen in other
objects (see Barlow \& Sargent\  1997; Hamann et al.\  1997).  The Ly$\alpha$\  and N$\:${\small V} lines are consistent with near zero coverage of the emission line gas, and complete coverage of the continuum, implying the cloud is larger than the continuum but smaller than the emission line region. Another possibility is that the absorption lines arise in several
regions which are sufficiently cool that they are unresolved,
saturated, and distributed over $\sim$ 80 km$\,$s$^{-1}$ in velocity
space so that at the instrument resolution used here the profile is
relatively smooth. We have not explored this far, but the conditions required 
for this to work at all for both H$\:${\small I} and N$\:${\small V} look somewhat contrived.

Savaglio \shortcite{sav98} observed the C$\:${\small IV} absorption for this system. It was noted that the equivalent width ratio for the C$\:${\small IV} doublet was lower than expected, further evidence for incomplete coverage of the source. The C$\:${\small IV} emission line is much weaker than the Ly$\alpha$\  emission line and the depth of the C$\:${\small IV} absorption lines require that the continuum is not completely covered.Therefore the absorption is shared between continuum and emission line. Assuming all absorption for all
ions involved comes from the same region uniformly and there is no
velocity-dependent or ion-dependent emission line region geometry, the best-fit covering factors are 50 per cent of the continuum source, and 75 per cent of the emission line region. This result is partly based on the low S/N EMMI spectrum, and further observations of the C$\:${\small IV}, Si$\:${\small IV} and O$\:${\small VI} absorption features are required to clarify the picture. 

\subsubsection{ $z=2.207$}

A blend of very broad (Doppler parameter $b \sim 200$ km$\,$s$^{-1}$) absorption lines, both for N$\:${\small V} $\lambda\lambda\, 1238, 1242  $ and Ly$\alpha$, were fitted for this system. The fit was complicated by uncertainty in the precise shape of the Ly$\alpha$\  \& N$\:${\small V} emission line continuum over the broad absorption feature, together with uncertainty in the covering factor. Also, the N$\:${\small V} $\lambda\, 1242  $ line from the $z=2.198$ system lay on top of the N$\:${\small V} $\lambda\, 1238  $ line from the $z=2.207$ system. We attempted to measure the covering factor, but no constraint was possible, so we assumed it to be 100 per cent. In order to obtain a satisfactory fit, however, we had to correct the continuum by subtracting a Gaussian centred on 3983\AA with FWHM = 670 km$\,$s$^{-1}$, and maximum depth 19 per cent of the original continuum level.

\section{Discussion}

Perhaps the most important enhancement of the HDF--S project over its predecessor is the inclusion of the STIS instrument observations of the field containing the $z=2.24$ QSO J2233-606. The STIS spectrograph will provide a high resolution (10 km$\,$s$^{-1}$) UV ($\lambda < 3500$\AA) spectrum of the QSO.  Furthermore, deep UV, optical and infra-red imaging of the STIS and surrounding flanking fields from both the HST and VLT observations will provide morphologies of galaxies near the QSO, and allow the accurate determination of photometric redshifts over the full range in absorber redshifts seen in the QSO
 spectrum.

One of the main aims of this project is to investigate the relationship between QSO absorbers and high redshift galaxies. The ground-based observations of the HDF--S QSO spectrum, reported here, and by Sealey et al. \shortcite{sdw98} and Savaglio \shortcite{sav98}, complement the STIS spectrum and are crucial to determining the properties of the intervening QSO absorbers and their association to galaxies seen in the STIS and flanking field images. In this paper we have reported eight heavy-element absorption systems which are potentially associated with galaxies. Combining this information with the imaging will allow investigation of the relationship between heavy element abundances and galaxy type, environment and impact parameter.

An associated absorption system of this quasar shows evidence for
incomplete coverage of both the emission-line region and the continuum
source. Further data are required to establish whether or not the
coverage is ion dependent, as it is in other such cases (e.g. Hamann et al.\  1997).

\section*{Acknowledgements}

\noindent   The observations were obtained with the Anglo-Australian Telescope. We thank S. Savaglio and H. Ferguson for making the EMMI and STIS data available. Much of the data reduction and analysis was perfomed on the Starlink-supported computer network at the Institute of Astronomy. P. Outram acknowledges support from PPARC.


\begin{thebibliography}{}
\bibitem[\protect\citename{Abraham et al.\ }%
 1996]{abr96} Abraham R. G., Tanvir N. R., Santiago B. X., Ellis R. S., Glazebrook K., van den Bergh S., 1996, MNRAS, 279, L47
\bibitem[\protect\citename{Barlow \& Sargent\ }%
 1997]{bar97} Barlow T.A., Sargent W.L.W., AJ, 113, 136
\bibitem[\protect\citename{Boyle\ }%
 1997]{boy97} Boyle B. J., 1997, AAO Newsletter, 83, 4
\bibitem[\protect\citename{Carswell et al.\ }%
 1992]{car92} Carswell R. F., Webb J. K., Cooke A. J., Irwin M. J., 1992, VPFIT Manual (program and manual available from rfc@ast.cam.ac.uk)
\bibitem[\protect\citename{Hamann et al.\ }%
 1997]{ham97} Hamann F., Barlow T.A., Junkkarinen V., Burbidge E.M., 1997, ApJ, 478, 80
\bibitem[\protect\citename{Kirkman \& Tytler\ }%
 1997]{kir97} Kirkman D., Tytler D., 1997, ApJ, 489, 123
\bibitem[\protect\citename{Lanzetta et al.\ }%
 1995]{lan95} Lanzetta K. M., Bowen D. V., Tytler D., Webb J. K., 1995, ApJ, 442, 538
\bibitem[\protect\citename{Lu et al.\ }%
 1996]{lu96} Lu L., Sargent W. L. W., Womble D. S., Takada-Hidai M., 1996, ApJ, 472, 509
\bibitem[\protect\citename{Madau et al.\ }%
 1996]{mad96} Madau P., Ferguson H. C., Dickinson M. E., Giavalisco M., Steidel C. C., Fruchter A., 1996, MNRAS, 283, 1388
\bibitem[\protect\citename{Outram et al.\ }%
 1998]{out98} Outram P. J., Chaffee F. H., Carswell R. F., 1998 (in preparation) 
\bibitem[\protect\citename{Savaglio\ }%
 1998]{sav98} Savaglio S., 1998, AJ (accepted)
\bibitem[\protect\citename{Sealey et al.\ }%
  1998]{sdw98} Sealey K. M., Drinkwater M. J., Webb J. K., 1998, ApJ, 499, L135
\bibitem[\protect\citename{Songaila\ }%
  1998]{son98} Songaila A., 1998, AJ, 115, 2184
\bibitem[\protect\citename{Steidel et al.\ }%
 1996]{ste96} Steidel C. C., Giavalisco M., Dickinson M., Adelberger K. L., 1996, AJ, 112, 352
\bibitem[\protect\citename{Webb\ }%
 1987]{web87} Webb J. K., 1987, Ph.D. thesis, Cambridge University
\bibitem[\protect\citename{Williams et al.\ }%
 1996]{wil96} Williams R. E. et al., 1996, AJ, 112, 1335
\end{thebibliography}
\end{document}